\documentclass[9pt,twocolumn,twoside]{optica}
\setboolean{shortarticle}{false}
\setboolean{minireview}{false}

\title{Fibers based on propagating bound states in the continuum}
\author[1,2]{Evgeny N. Bulgakov}
\author[1,*]{Almas F. Sadreev}

\affil[1]{$^1$ Kirensky Institute of Physics, Federal Research Center KSC SB RAS, 660036
Krasnoyarsk, Russia}
\affil[2]{$^2$Siberian State Aerospace University, Krasnoyarsk 660014, Russia}

\affil[*]{Corresponding author: almas@tnp.krasn.ru}

\ociscodes{(060.2420) Fibers, polarization-maintaining; (060.3735) Fiber Bragg gratings;
(230.7400)Waveguides.}

\begin{abstract}
We show that a circular periodic array of $N$ dielectric cylinders
supports nearly bound states in the continuum (BICs) propagating
along the cylinders. These propagating nearly BICs with extremely
large $Q$ factors of order $exp(\lambda N)$ are surrounded by
resonant modes weakly leaking into the radiation continuum. We
present leaky zones in the vicinity of different types of BICs:
symmetry protected nearly BICs with the resonant width
proportional to the squared propagation constant $\Gamma \sim
k_z^2$, non-symmetry protected nearly BICs with finite propagation
constant $k_c$ with $\Gamma\sim (k_z-k_c)^2$ and non-symmetry
protected nearly BICs with $\Gamma\sim k_z^4$. The latter
propagating nearly BICs  can serve for transmission of
electromagnetic signal paving a way to  novel type of optical
fibers. We also demonstrate weakly leaking resonant modes which
carry orbital angular momentum.

\end{abstract}

\setboolean{displaycopyright}{true}

\begin{document}

\maketitle

\section{Introduction}
Standard optical fibers guide light using total internal reflection. This restricts their
optical properties, because only solid or liquid materials can be used for the fiber core.
There are no suitable cladding materials which have a sufficiently low refractive index to confine
light by total internal reflection in a vacuum or a gas core.

Substantial efforts have been invested over the past years in
fabricating photonic crystals materials that have a periodic
modulation of the refractive index on the scale of the optical
wavelength. The interest in such materials lies in their ability
to strongly reflect light of certain frequencies. For example,
structure consisted of periodically designed layers forms
one-dimensional photonic crystal (PhC) which exhibit band gaps at
optical frequencies (photonic band gaps) \cite{Joan}. Light that
is incident upon a band-gap material from the outside would be
totally reflected. Similarly, light that existed at a
structural-defect site in such a material would be permanently
trapped, being unable to propagate through the lattice. Photonic
band gap (PBG) structures offer the opportunity to design new
optical properties into existing materials by wavelength-scale
periodic micro structuring of the material morphology
\cite{KnightScience}. One can imagine that such a structure of
order of ten layers can be rolled up to form cladding capable to
almost perfectly trap light inside realizing fiber. Another design
of two-dimensionally periodic structures in the form of long, fine
silica fibers that have a regular array of tiny air holes running
down their length constitute artificial two-dimensional PhC with
lattice constants on the order of micrometers \cite{KnightOL}.

However, the demand of perfectness of such fibers enormously
enlarges their cross-section. In the present paper we propose a
different design of fibers based on the capability of a periodic
array of dielectric cylinders to trap light at certain
frequencies. The property is based on a fundamental family of
localized solutions of Maxwell's equations, so called bound states
in the continuum (BICs). Recently BICs with zero Bloch vector were
reported in infinitely long periodic arrays of dielectric
cylinders \cite{Shipman,Marinica,HsuNature,Weimann,
Wei,PRA2014,BoZhen,Yang,Foley,Hu&Lu,Song,Yuan,ZWang,Sadrieva,AdvEM,YuanOL,Hu&Lu2017}.
BICs propagating along the array were also shown to exist
\cite{PRA2014,Zou,B&MOL,Yuan&Lu,Hu&LuJOSAB,Yuan&Lu95,Zhen,B&MPRL,B&M2017}.
Our primary interest is in the BICs which can propagate along the
cylinders \cite{Zhen,B&M2017} giving rise to a new family of
guided modes with frequencies above the light line.

Physically, the occurrence of BICs in the infinite array of
cylinders is the result of the periodicity of the array that
quantizes the radiation continua in the form of diffraction
continua \cite{PRA2014,AdvEM}. Obviously, the infinite array of
dielectric cylinders is an unrealistic limit. In practice we deal
with finite number $N$ of cylinders which have material losses
given by the imaginary part of refractive index, structural
fluctuations of cylinders, the effect of substrate {\it etc},
transforming the ideal BIC into a resonant mode with small
resonant width \cite{Song,AdvEM,Ni,Sadrieva,B&MOE}. Although the
full range study of these factors  is still far from completion it
was shown that the $Q$ factor of the symmetry protected quasi BICs
grows quadratically with $N$.
%--------------------------------------------------------------------------------------------Fig.1
\begin{figure}[t]
\vskip 0.05in
\begin{center}
% \setlength{\epsfxsize}{3.5in}
% \centerline{\epsfbox{cascade.eps}}
\includegraphics[width=10cm,clip=]{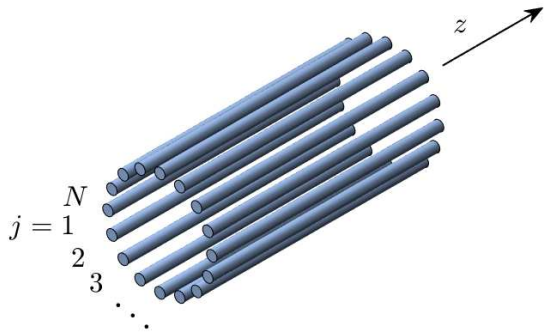}
% \vskip 0.1in
\caption{$N$ infinitely long circular dielectric cylinders with
radius $a$ stacked  parallel to each other in a circle of radius $R$.}
\label{fig1}
\end{center}
\vskip -0.2in
\end{figure}

However if the array of cylinders is rolled into a circle as shown
in Fig. \ref{fig1} the $Q$ factor grows exponentially with $N$
\cite{Lu&Liu,PRA97}. In practice such Q-factors make the nearly
BICs in the circular array indistinguishable from true BICs in the
infinite array of cylinders \cite{PRA97}.

In the present paper we demonstrate a few examples of the nearly
BICs surrounded by propagating resonant modes weakly leaking into
the radiation continuum. The property of the nearly BICs to serve
as modes guided above the light line paves a way to new designs of
fibers composed of $N$ dielectric cylinders circularly packed
parallel to each other. These nearly BICs fill the core of the
fiber and can carry  orbital angular momentum (OAM) $m$. Each type
of the above listed nearly BICs is hosted by a leaky zone with
high $Q$ surrounding the nearly BIC.

%%%%%%%%%%%%%%%%%%%%%%%%%%%%%%%%%%%%%%%%%%%%%%%%%%%%%%%%%%%%%%%%%%%%%%%%%%%%%%%%%
\section{Near BICs propagating along the fiber}
\begin{figure}[t]
\vskip 0.05in
\begin{center}
% \setlength{\epsfxsize}{3.5in}
% \centerline{\epsfbox{cascade.eps}}
\includegraphics[width=10cm, clip=]{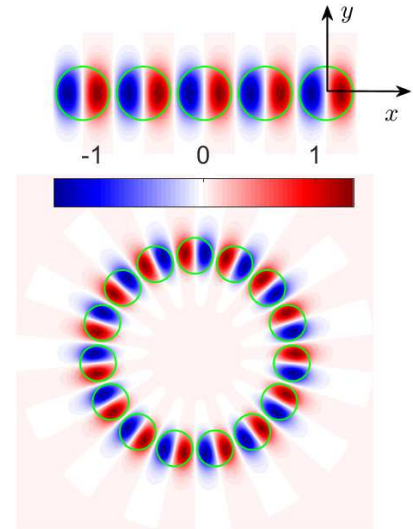}
% \vskip 0.1in
\caption{z-component of the electric field of the symmetry
protected standing wave BIC in the linear periodic array of
dielectric cylinders with $a=0.44$ and $\epsilon=15$ with
frequency $k_{0c}=1.8315$ and its counterpart, nearly standing BIC
in the circular array of the radius $R=7.25a$ of 15 cylinders with
frequency $k_{0c}=1.837$ with $m=0$ from Ref \cite{PRA97}. The
radius of the cylinders $a$ is given in terms of the azimuthal
period of circular array $h=2\pi R/N$ where $R$ is the radius of
circular array. Respectively the wave numbers are given in terms
of inverse $h$.}
\label{auxil}
\end{center}
\vskip -0.2in
\end{figure}

%%%%%%%%%%%%%%%%%%%%%%%%%%%%%%%%%%%%%%%%%%%%%%%%%%%%%%%%%%%%%%%%%%%%%%%%%%%%%%%%%
Here for brevity  we omit the details of calculations. The
calculations are based on the theory of scattering by a finite
cluster of cylinders \cite{Yasumoto,PRA97}. We start  with the
simplest symmetry protected standing wave nearly BIC whose
coupling with the radiation continuum is exponentially weakened
because of symmetry incompatibility \cite{Lu&Liu,PRA97}. This
nearly BIC as shown in Fig. \ref{auxil} originates from a true
standing wave BIC in the infinite periodic array of dielectric
cylinders at the $\Gamma$-point first reported by Shipman and
Venakides \cite{Shipman}.
%--------------------------------------------------------------------------------------------Fig.2
The electric field of the BIC solution directed along the
cylinders is even relative to the direction perpendicular to the
plane of the array (y-axis in Fig. \ref{auxil}) and odd relative
to $x\rightarrow -x$ where $x$ and $y$ are local coordinate system
tied to the center of cylinder. The z-component of magnetic field
of the symmetry protected BIC equals zero to define the nearly BIC
as E-polarized.  When $N$ cylinders are rolled up into a circle
still the symmetry of the solution relative to $x\rightarrow -x$
plays the key role to provide extremely small coupling with the
radiation continuum in the form of outgoing cylindrical waves for
$k_z=0$ and $k_m=0$ where $k_m=2\pi m/N=0$. The integer $m=0, \pm
1, \pm 2, \ldots, \pm (N-1)$  specifies OAM. For the infinite
periodic array of cylinders $k_m$ would specify the Bloch wave
number along the array.

\begin{figure}[t]
\vskip 0.05in
\begin{center}
% \setlength{\epsfxsize}{3.5in}
% \centerline{\epsfbox{cascade.eps}}
\includegraphics[width=8cm, clip=]{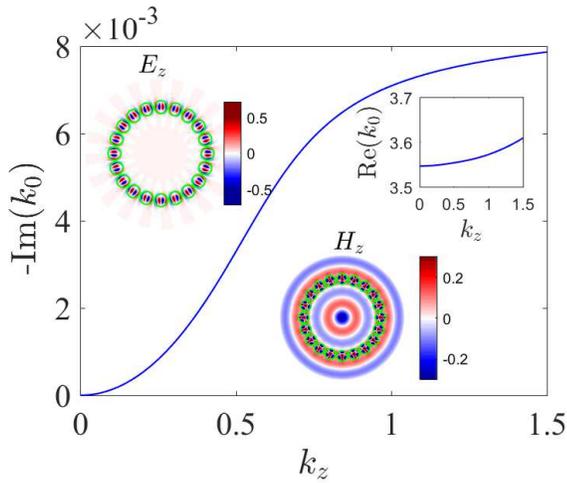}
% \vskip 0.1in
\caption{Leaky
zone of resonant modes propagating along the fiber consisted of 20
silicon cylinders with $\epsilon=15$ and radius $a=0.44=0.1382R$.
Insets show z-components of electric and magnetic fields at
$k_0=3.6086-0.007866i, k_z=1.5, m=0$.}
\label{fig2}
\end{center}
\vskip -0.2in
\end{figure}

The radius of the cylinders $a$ is given in terms of the azimuthal
period of circular array $h=2\pi R/N$ where $R$ is the radius of
circle. Respectively, the wave numbers are given in terms of
inverse $h$. The dispersion curves are computed by solving the
dispersion equation $f(k_0,k_z)=0$ through analytical continuation
of $k_0$ into complex plane, where $k_0=\omega h/c$ is the vacuum
wave number, and $k_z$ is the propagating constant, the wave
number  along the cylinders. Fig. \ref{fig2} shows the real and
imaginary parts of complex eigenfrequencies for the case of 20
silicon cylinders. The resonant width and frequency depend on
$k_z$ quadratically for small $k_z$ as seen from Fig. \ref{fig2}.
Such a behavior is typical for the guided modes in the vicinity of
the $\Gamma$-point in infinite arrays
\cite{Yuan&Lu95,B&MPRL,B&MOE,Polishchuk}. The $Q$ factor of the
eigenmode is given by equation $Q=-{\rm Re}(k_0)/2{\rm Im}(k_0)$.
Insets in Fig. \ref{fig2} show profiles of electromagnetic fields
(z-components of electric and magnetic field) at $k_z=1.5$. This
mode converts into a standing wave E-polarized nearly BIC with
$H_z=0$ at $k_z=0$ (see Fig. 5 in Ref. \cite{PRA97}).
%--------------------------------------------------------------------------------------------Fig.3
Therefore when the wave number $k_z$ moves away from zero
not only the $Q$ factor reduces but also both
polarizations are mixed as seen from insets in Fig. \ref{fig2}.
One can see that magnetic field fills whole inner space of the
fiber as different from the electric field which mostly localized
inside the cylinders. That is related to that the electric field
is odd relative to $x\rightarrow -x$ to be mostly localized around
the cylinders while the magnetic field is even to fill whole inner
space of the fiber \cite{PRA97}.

\begin{figure}[t]
\vskip 0.05in
\begin{center}
% \setlength{\epsfxsize}{3.5in}
% \centerline{\epsfbox{cascade.eps}}
\includegraphics[width=8cm,clip=]{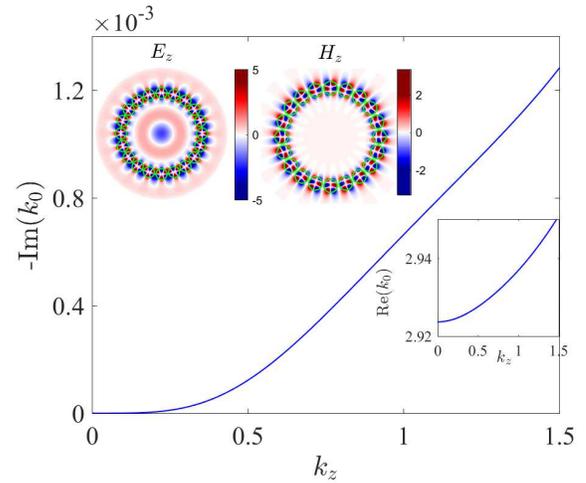}
% \vskip 0.1in
\caption{Leaky zone of resonant mode of the circular array of
cylinders with $a=0.43084=0.1354R$ which converts into the non
symmetry protected E-polarized nearly BIC at $k_z=0$ and
$a=0.43084=0.1354R$ Insets show z-components of electric and
magnetic fields at $k_0=2.9508-0.00128i, m=0, k_z=1.5$.}
\label{fig3}
\end{center}
\vskip -0.2in
\end{figure}

The Fig. \ref{fig3} shows the dispersion curve and the resonant
width of the mode which originates from the non-symmetry protected
standing  E-polarized nearly BIC at $k_z=0$.
%Similar to  the former symmetry
%protected nearly BIC in Fig. \ref{fig2} this nearly BIC is the H-polarized.
This nearly BIC is symmetry protected in respect to magnetic field
and due to tuning the cylinder radius acquires exponentially small
coupling with the radiation continuum in respect to the electric
field to achieve $Q=2.6\cdot 10^8$. When $k_z$ moves away from
zero the resonant mode mixes both polarizations. For the infinite
array of cylinders electromagnetic field of this non-symmetry
protected BIC were localized around the cylinders. In the circular
array the leaky mode has even electric field filling whole core of
the fiber, while the odd magnetic field remains localized around
the cylinders as shown in insets in Fig. \ref{fig3}.
%--------------------------------------------------------------------------------------------Fig.4

\begin{figure}[t]
\vskip 0.05in
\begin{center}
% \setlength{\epsfxsize}{3.5in}
% \centerline{\epsfbox{cascade.eps}}
\includegraphics[width=8cm, clip=]{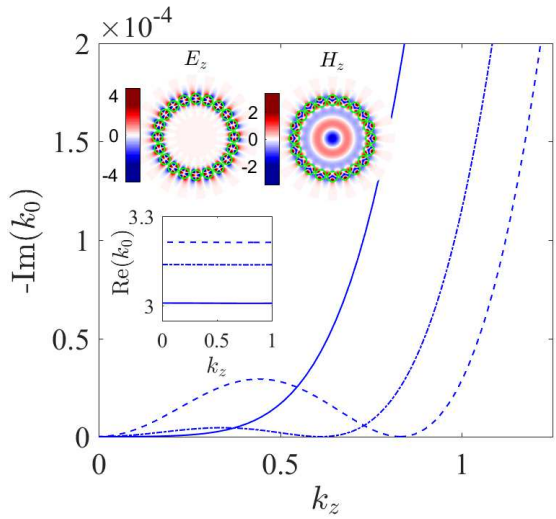}
% \vskip 0.1in
\caption{Leaky zone of resonant mode at $a=0.418=0.1313R$ (dash
lines) with two nearly BICs at $m=0$ and $k_z=0$ and $k_z=0.83$,
at $a=0.43=0.1351R$ with two nearly BICs at $k_z=0$ and
$k_z=0.605$ which finally collapses into the mode with single
nearly BIC at $k_z=0$ whose resonant width has an asymptote
$k_z^4$ at $a=0.453=0.1382R$ (solid lines). Insets show
z-components of electric and magnetic fields at
$k_0=3.2124-3.95\cdot 10^{-8}i, k_z=0.83$ for $a=0.418$.}
\label{fig4}
\end{center}
\vskip -0.2in
\end{figure}

The most interesting feature of this nearly BIC, however, is a
quaternary dependence of the resonant width on $k_z$ as shown Fig.
\ref{fig3}. Such a behavior of the resonant width was shown in
Ref. \cite{B&M2017} relative to $k_z$ and in Refs.
\cite{B&MOE,Yuan&Lu2018} relative to the Bloch wave number along
the infinite periodic array. Above  we considered the leaky zones
of the resonant modes in Figs. \ref{fig2} and \ref{fig3}
originated from standing wave nearly BICs which are suitable for
signal transmission along the fiber because of slow velocity of
the signal. The Fig. \ref{fig4} shows the leaky zones which holds
BIC point $k_{zc}\neq 0$. One can see the  evolution of the
resonant width vs the propagation constant with increasing of the
cylinder's radius $a$. For the first two choices $a=0.418$ and
$a=0.43$ there are two points where the resonant width nearly
turns to zero (dash and dash-dot lines in Fig. \ref{fig4}). The
first point $k_{zc}=0$ corresponds to the symmetry protected
E-polarized standing wave nearly BIC with $Q=1.6\cdot 10^7$ for
$a=0.418$ and $Q=5\cdot 10^7$ for $a=0.43$. The second point
corresponds to the propagating nearly BIC with mixed
polarizations. The propagation constant $k_{zc}$ turns to zero
with the  increase of cylinder's radius with the two BICs
coalescing at $a=0.453$ at $k_{zc}=0$. The leaky resonant modes
hosting this standing wave nearly BIC at the point of coalescence
acquires quaternary dependence of the resonant width $-{\rm
Im}(k_0)\sim k_z^4$ as shown in Fig. \ref{fig4} by solid line.
That phenomenon was studied in details for the case of the
infinite array of cylinders and spheres in Refs.
\cite{B&MPRL,Yuan&Lu} as a result of preservation of topological
charge in two-dimensional space of two polarizations of the BIC.
It is remarkable for  $k_z$ in the wide range the resonant width
is smaller than $2.5\cdot 10^{-4}$ as shown in Fig. \ref{fig4} by
dash and dash-dot line.  A weak dependence of the $Q$ factor on
the wave number allows to use this nearly BIC for signal
transmission with high efficiency.
%--------------------------------------------------------------------------------------------Fig.5

\begin{figure}[t]
%\vskip 0.05in
\begin{center}
% \setlength{\epsfxsize}{3.5in}
% \centerline{\epsfbox{cascade.eps}}
\includegraphics[width=9cm,clip=]{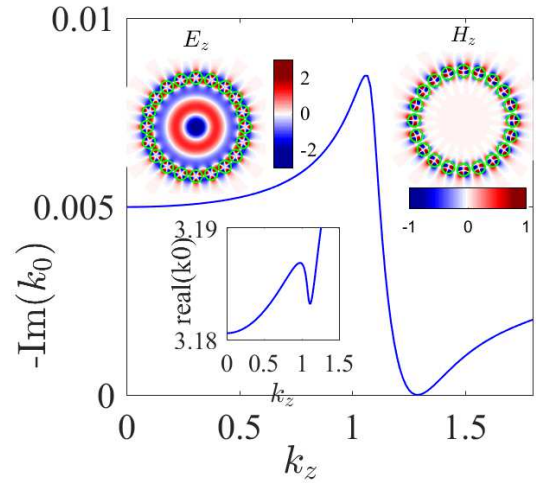}
% \vskip 0.1in
\caption{Leaky zone of resonant mode which converts into the
symmetry protected nearly BIC for $a=0.4=0.1257R$. Insets show
z-components of electric and magnetic fields at the complex
eigenvalue $k_0=3.1916-5\cdot 10^{-9}i$ and the wave number
$k_z=1.29, m=0$.}
\label{fig5}
\end{center}
\vskip -0.2in
\end{figure}

Fig. \ref{fig5} shows the resonant mode which holds only the
propagating nearly BIC at finite values of the wave number but not
a standing wave nearly BIC.
%--------------------------------------------------------------------------------------------Fig.6

This nearly BIC has mixed polarizations  with the even electric
field $E_z$ filling whole core of the fiber and the odd magnetic
field localized in the vicinity of cylinders. The mode has $Q$
factor of order $3\cdot 10^2$ at $k_z=0$. The $Q$ factor decreases
when the wave number goes away from zero but then again goes to
extremely large value $3.2\cdot10^8$ when $k_z$ reaches $k_z=1.29$
as shown in Fig. \ref{fig5}. The Dispersion curve shows a
non-monotonic behavior that is related to an avoided crossing of
two neighboring resonances.

\begin{figure}[t]
\vskip 0.05in
\begin{center}
% \setlength{\epsfxsize}{3.5in}
% \centerline{\epsfbox{cascade.eps}}
\includegraphics[width=8cm,clip=]{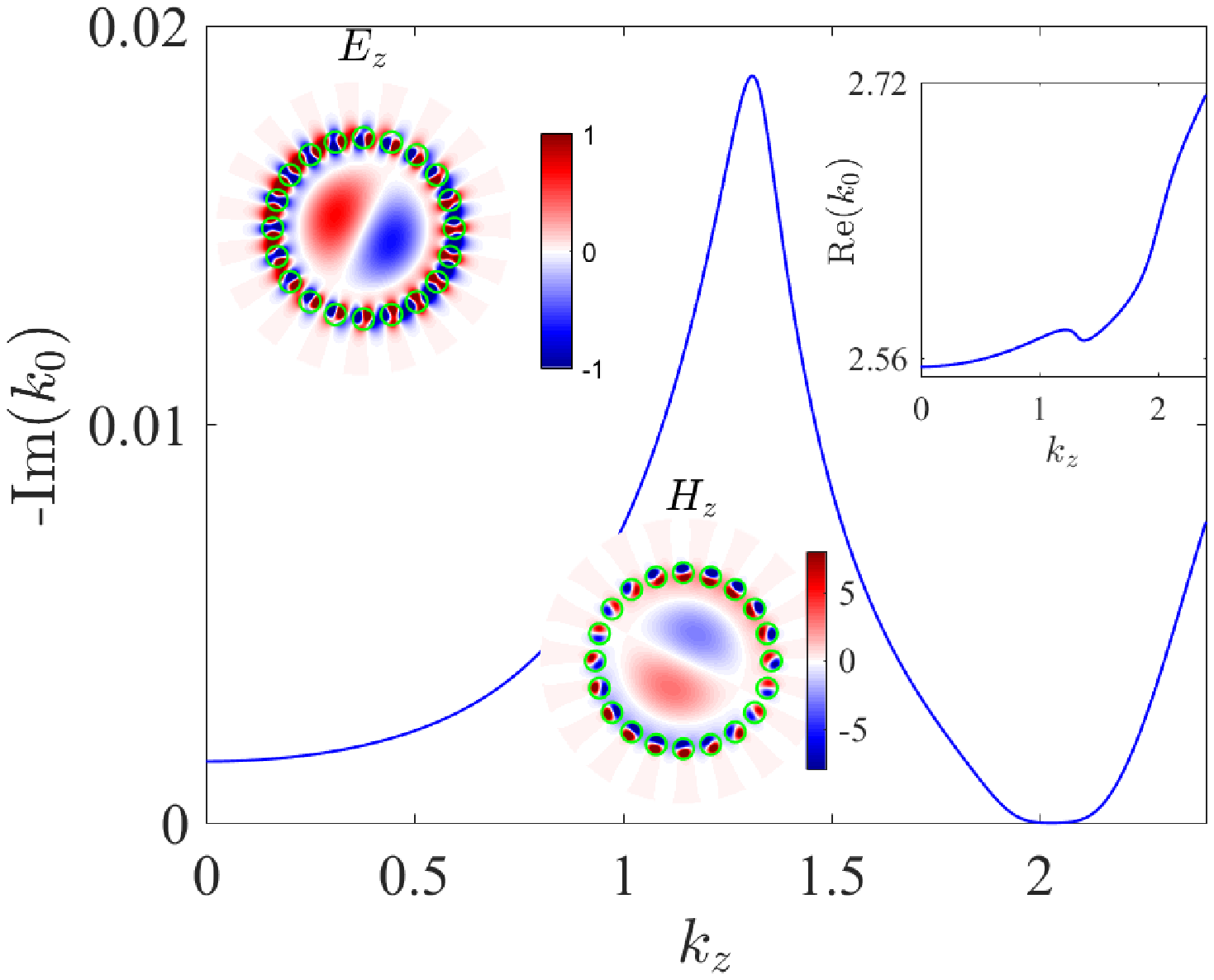}
% \vskip 0.1in
\caption{Resonant mode which converts into the nearly BIC with OAM $m=1$
for $a=0.369=0.116R$. Insets show z-component of the electric
field of the propagating nearly BIC at $k_0=2.644-5\cdot10^{-9}i,
k_z=2.027$.}
\label{fig6}
\end{center}
\vskip -0.2in
\end{figure}

The former cases with zero OAM do not need tuning the cylinder
radius. Once $m\neq 0$ the propagating wave nearly BICs with
nonzero OAM need tuning  the radius as it was shown in Ref.
\cite{PRA97}. These propagating wave nearly BICs with OAM are
shown in Figs. \ref{fig5} ($m=1$) and \ref{fig6} ($m=3$).
%--------------------------------------------------------------------------------------------Fig.7

%--------------------------------------------------------------------------------------------Fig.8
\begin{figure}[t]
\vskip 0.05in
\begin{center}
% \setlength{\epsfxsize}{3.5in}
% \centerline{\epsfbox{cascade.eps}}
\includegraphics[width=8cm,clip=]{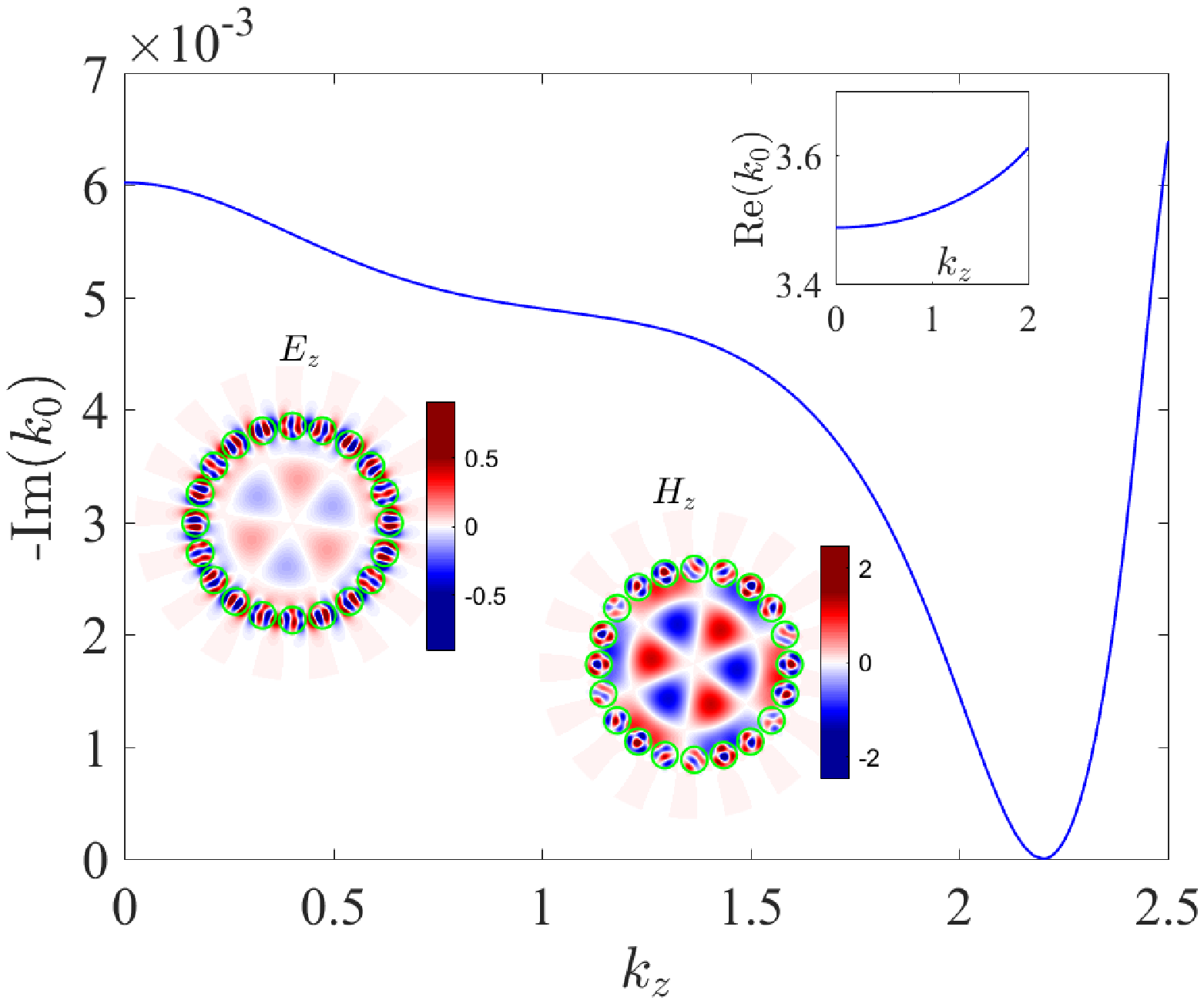}
% \vskip 0.1in
\caption{Leaky zone of resonant mode which converts into the
nearly BIC with OAM $m=3$ for $a=0.4327=0.136R$. Insets show
z-component of electric and magnetic fields of the propagating
nearly BIC  at $k_0=3.6544-2\cdot10^{-7}i, k_z=2.2$.}
\label{fig7}
\end{center}
\vskip -0.2in
\end{figure}

\section{Discussion and conclusions}
\begin{figure}[t]
\vskip 0.05in
\begin{center}
% \setlength{\epsfxsize}{3.5in}
% \centerline{\epsfbox{cascade.eps}}
\includegraphics[width=9cm,clip=]{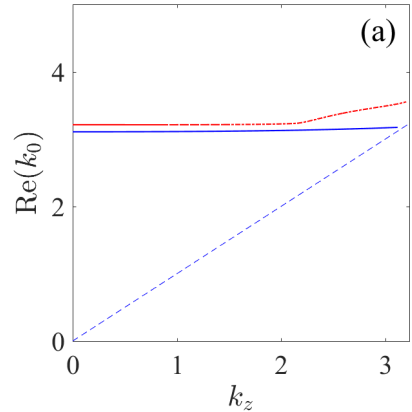}
\includegraphics[width=9cm,clip=]{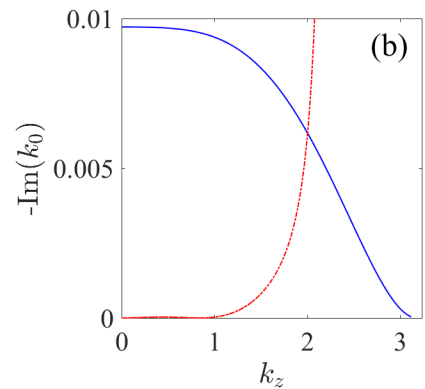}
% \vskip 0.1in
\caption{Dispersion (a) and resonant width (b) of guided mode in isolated dielectric cylinder
with $a=0.418$ and $m=0$ (solid lines) compared to the case shown in Fig. \ref{fig4}
(red dash-dot lines). Thin dash line shows light line.}
\label{fig8}
\end{center}
\vskip -0.2in
\end{figure}

First, it is interesting to compare propagating nearly BICs in the
circular array of cylinders with guided modes propagating along an
isolated dielectric cylinder \cite{Jackson}. That comparison is
given in Fig. \ref{fig8}. One can see that the frequencies as
dependent on the propagation constant $k_z$ behave very similar to
one another while the resonant widths  are strikingly  different.
If the frequency of the guided mode in the isolated cylinder is
below the line of light the mode can propagate along the cylinder
without leakage. As soon as the frequency is above the line the
mode becomes leaky  as shown in Fig. \ref{fig8} (b) by solid line
while the widths of  the nearly BICs including the resonant modes
surrounded the BIC have extremely small leakage above the light
line in a rather wide domain of the propagation constant.

%--------------------------------------------------------------------------------------------Fig.9

Apparently, the choice of the fiber of $N$ dielectric cylinders of
circular cross-section is not the best with the technology view
point. In general there can be any circular dielectric structure
which possess a symmetry relative to azimuthal discrete rotations
$\phi\rightarrow \phi+2\pi n/N$ where $n=1, 2, 3, \ldots , N$ and
$N$ is an integer. In practice the fiber can be chosen in the form
of single dielectric cylinder with periodical grating on its
surface. The present type fiber composed of $N$ dielectric
cylinders has a unique property to exponentially enlarge the $Q$
factor with $N$ \cite{Lu&Liu,PRA97} for specific solutions, i.e.,
that is nearly BICs.  These solutions are localized within the
fiber in spite that the frequency of the solution is embedded into
the radiation continuum. The fiber can support various nearly BICs
mostly standing waves. These BICs are surrounded by weakly leaking
resonant modes with the $Q$ factor proportional to inverse of
$k_z^2$. The symmetry protected nearly BICs do not need tuning of
the cylinder radius which makes them interesting with the
technological point of view. There are also non-symmetry protected
nearly BICs which occur via tuning the cylinder radius. The
resonant modes surrounding these nearly BICs have extremely weak
quaternion dependence of the resonant width on the propagation
constant to be specially  interesting for signal transmission. The
non-symmetry protected propagating nearly BICs surrounded by
resonant modes with $Q$ factor inversely proportional to
$(k_z-k_{zc})^2$ are the most interesting for signal processing in
the fiber. They do not need tuning cylinder radius in contrast to
nearly BICs which carry OAM.

It is clear that transmission of electromagnetic signals over the
fiber requires some finite range of frequencies. Because of
discreteness of the BIC frequency propagation of signals will be
accompanied by leakage. However the majority of resonant widths do
not exceed one percent of the frequency. The propagation length
which is given by the decay rate of nearly BICs into the radiation
continuum. Its value can be accessed as \cite{B&MOL}
\begin{equation}\label{Lz}
    \frac{L}{\lambda}=\left |\frac{d{\rm Re}(k_0)}{dk_z}\right |
    \frac{k_0}{{2\pi \rm Im}(k_0)}.
\end{equation}
For example we obtain $L/\lambda \approx 10^4$ for $a=0.418$ (see
Fig. \ref{fig4}) and $L/\lambda \approx 5\cdot 10^6$ for $a=0.4$
(see Fig. \ref{fig5}) where $\lambda$ is the wavelength. Therefore
the propagating nearly BICs  can serve for propagation of
electromagnetic signals with high quality. That prompts to use the
circular array of cylinders as a novel type of optical fibers.

% $0.01$

\section*{Funding Information}
This work was partially supported by Ministry  of  Education  and
Science  of  Russian Federation (State contract  N 3.1845.2017)
and the RFBR Grants No.16-02-00314 and No.17-52-45072.

\section*{Acknowledgments}
The authors thank D.N. Maksimov  for assistance and discussions.

\newpage

% Bibliography
%\bibliography{sample}

% Full bibliography will be added automatically on a new page for Optics Letters submissions. This command is ignored for journal article submissions.
% Note that this extra page will not count against page length.
%\bibliographyfullrefs{sample}
%Manual citation list

\end{document}